# Anomalous Phonon Renormalization in Single Crystal of Silicon

Himanshi, Birender Singh and Pradeep Kumar

*School of Basic Sciences, Indian Institute of Technology Mandi, Mandi-175005, India*

**Abstract.** The temperature dependence of the first-order phonon mode of single crystal of Silicon (Si) is determined by Raman scattering in a broad temperature range of 4-623 K. Our studies reveal the anomalous red-shift of the Raman active phonon mode at temperature (~ 50 K) attributed to the anomalous expansion of Si in the low temperature region. Silicon shows negative thermal expansion below 120 K, however, odd behaviour is also observed at very low temperatures i.e., softening of the Si crystal is detected below 40 K. This peculiar behaviour of Si is described by the anomalous phonon anharmonicity observed at low temperature.

## INTRODUCTION

Conventionally, on heating, most solids show positive thermal expansion i.e. they expand upon heating. However, a few solids are known to show contraction with the increase in temperature. This phenomenon is known as negative thermal expansion and occurs in a wide variety of solids e.g., $ZrW_2O_8$, quartz, silicon, and many zeolites [1-4] over varying temperature ranges. Materials exhibiting negative thermal expansion can be used to compensate for the undesired positive expansion of other materials, creating ceramics, composites or devices with controllable overall negative, zero or positive coefficient of thermal expansion herein given by α. Thus, the temperature dependence of such materials is being paid a great deal of attention in research and development. Si is stronger than stainless steel, has a thermal conductivity about half that of aluminium, is transparent too much of the infrared radiation spectrum, and can form a stable oxide. Si also has a thermal expansion coefficient that is four times lower than titanium and about an order-of-magnitude lower than aluminium and stainless steel [5]. Therefore, Si with such assorted properties has been widely used in various fields of science and technology such as in electronic devices and tech products. Also, the unique properties of Si enable it to become most of the mass of a satellite and it can simultaneously function as structure, heat transfer system, radiation shield, optic, and a semiconductor substrate. It also supports the exterior of a space capsule that experiences high and low temperatures variations [5]. Furthermore, the thermal properties of Si are of importance for silicon-based electronics, nano-mechanics, photovoltaics, thermoelectrics, and batteries.

However, despite the extensive use of Si in aerospace and tech industry its thermal expansion behaviour at low temperatures is still obscure. Negative thermal expansion of Si at low temperatures has been reported in earlier studies, in terms of lattice parameters, thermal expansion coefficients and grüneisen parameters by X-Ray diffraction [3], inelastic neutron scattering [6], and by developing theoretical model [7]. D.S. Kim et al. [6] have acknowledged the odd thermal expansion behaviour of Si at low temperatures by inelastic neutron scattering but to our knowledge, there has not been a comprehensive study on Si at cryogenic temperatures via Raman spectroscopy. Raman scattering is a non-destructive analytical method applied to study on a molecular or sub-molecular scale of molecules or atom assemblies in any physical state, at any temperature or pressure. For a solid material, this technique allows identification of the characteristic vibrations of its functional groups or their collective oscillations termed as phonons. The temperature dependent Raman scattering has been successfully measured in studies on Si, Ge, diamond [8], and GaAs, etc. and are interpreted in terms of anharmonic processes which leads to a better understanding of electronic properties of semiconductors at different temperatures.

In present work, we have used Raman scattering as an optical probe for characterizing the effect of temperature on a single crystal of Si and report the temperature dependence of the frequency shift and line-width of the first-order phonon mode near ~520 $cm^{-1}$ in a wide temperature range starting from 4 K to 623 K. We note that the thermal expansion of Si over the temperature range from 4 - 800 K estimated using a model has shown a different course and it has also been suggested that the thermal expansion coefficient of Si behaves differently in three temperature regions: positive for temperature below 15 K, negative between 15 and 125 K, and positive again above 125 K till 800 K [9].

# RESULTS AND DISCUSSION

The Raman spectra were obtained using the 532nm line of a diode laser. Unpolarised temperature dependent Raman measurements were carried out with Horiba Labram HR evolution Raman Spectrometer in back scattering geometry using confocal optics at different temperatures. The laser power targeted at the sample was estimated to be less than 2mW, to avoid any heating effect on the sample. The scattered light was detected using 1800 grating/mm coupled with Peltier cooled charge coupled device (CCD) detector. The experiment was carried out over a wide temperature range from 4 K to 623 K, with a temperature accuracy of ±0.1K.

The spectra at low temperatures were recorded in close cycle refrigerator using liquid helium as the coolant. For heating, the sample from 340 K to 623 K a Linkam temperature stage with a quartz window was used. For each temperature measurement, the temperature was stabilized for about 10 minutes.

Jayant S.Shah & M.E. Shraumanis [3] have studied the thermal expansion behaviour of Si from 180-40 K by the X-Ray diffraction method. They found that the coefficient of thermal expansion of Si at 120 K is zero and negative below this temperature. However, the anomalous thermal expansion behaviour below 40 K has not studied broadly and is demonstrated in our experiment. Figure 1 (a and b) shows Raman spectra of single crystal of Si recorded at different temperatures in the temperature range of 4 K to 623 K and Raman spectra at 4 K fitted with a Lorentzian function, respectively. In order to have quantitative information about the self-energy parameters i.e. mode frequency and line-width, we have fitted all recorded spectra at different temperature with Lorentzian function.

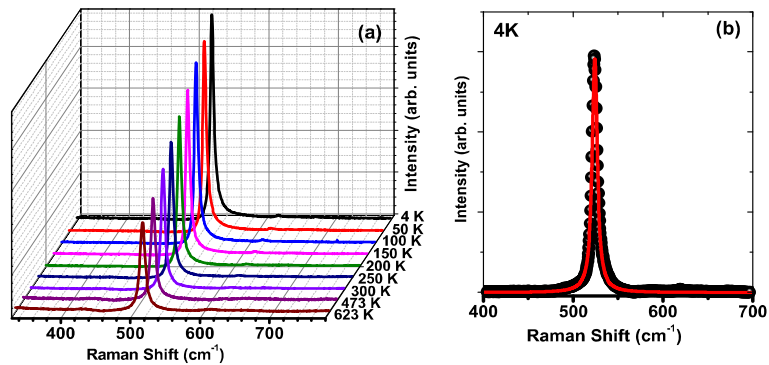

**FIGURE 1.** (a) Temperature evolution of the Raman spectra of single crystal of Si from 4 K to 623 K and (b) Raman spectra recorded at 4 K, solid thick red line shows Lorentzian function fit to the experimental data.

Figure 2(a) illustrate the frequency shift of the Raman active phonon mode of Si as a function of temperature from 4 K-623 K. The phonon mode frequency is observed to exhibit different behaviour in different temperature ranges. Temperature dependence of the mode frequency above 60 K, may be understood within the anharmonic picture i.e., phonon frequency observed from 623 K to ~ 60 K increases with decrease in temperature. However,

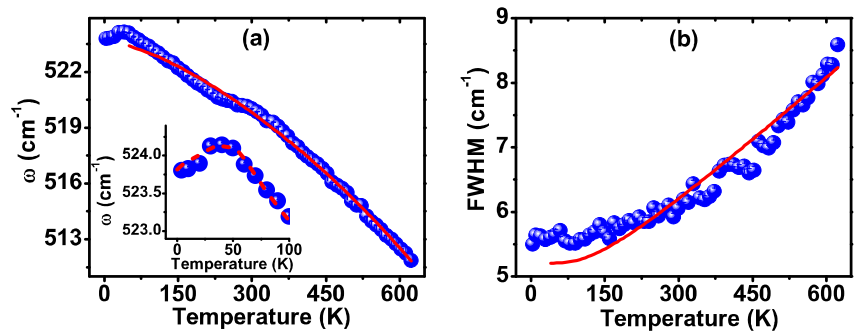

**FIGURE 1.** (a) Temperature evolution of the mode frequency and the inset shows the frequency as a function of temperature in small temperature window clearly reflecting the softening below 50 K, dotted line is guide to eye. (b) FWHM as a function of temperature. Solid lines are the fitted curve described in the text.

the anomalous softening of the mode frequency observed at lower temperature (below ~ 50-60 K) cannot be explained by this picture. This anomalous softening of the mode frequency below ~50-60 K suggests the expansion of the Si crystal instead of the monotonic contraction below this temperature. Within the simplistic picture a solid may be assumed to be made of atoms connected with the springs and under normal circumstances with decrease in temperature a solid contract and as a result spring constant ($k$) of the spring connecting two atoms also increases. As the atomic vibrational frequency (ω) is directly proportional to $\sqrt{k}$. So, with increase in k with decrease in temperature, ω should increase. However, if there is an anomalous expansion of a solid with decrease in temperature then that will lead to decrease in spring constant and ultimately red shift of the phonon frequency. Our observation of the anomalous softening of the mode frequency below ~50-60 K clearly suggests the anomalous expansion of Si below this temperature. With change in the temperature renormalization of the phonon mode frequency mainly arise from two factors i.e. (i) thermal expansion of the lattice ($\Delta\omega_E(T)$) (ii) anharmonic effect ($\Delta\omega_A(T)$). The change in the phonon mode frequency considering the above effect may be given as $\Delta\omega(T) = \Delta\omega_E(T) + \Delta\omega_A(T)$. $\Delta\omega_E(T)$ is given as

$\Delta\omega_E = \omega_0[\exp(-3\gamma_0\int_0^T \alpha(T)dT) - 1]$, and $\gamma_0$ is the Grüneisen parameter taken as 2 for Silicon, and $\alpha(T)$ is linear thermal expansion coefficient of the material and the product of $\gamma_0$ and $\alpha(T)$ may be written as $3\gamma_0\alpha(T) = a_0 + a_1 T + a_2 T^2 + ...$ where, $a_0$, $a_1$ and $a_2$ are constant. $\Delta\omega_A(T)$ is given as $\Delta\omega_A = c[1+(2/(e^x -1))]$ where, $x = \hbar\omega_0/2k_B T$ ($k_B$ is the Boltzmann constant), $\omega_0$ is mode frequency at $T$ = 0 K and $c$ is a constant. To extract the thermal expansion coefficient above 60 K, we have fitted the mode frequency using the following equation [10-12]:

$$\omega(T) = c(1+(2/(e^x -1))) + \omega_0\exp(-3\gamma_0\int_0^T \alpha(T)dT) \quad \ldots\ldots (1)$$

The extracted thermal expansion coefficient above ~ 60 K is shown in Figure 3, and the temperature evolution of $\alpha(T)$ matches quite well with the other experimental studies [7, 9].

Figure 2(b) shows the dependence of full width at half maximum (FWHM) of the phonon mode as a function of temperature. The FWHM exhibit normal temperature dependence behaviour i.e. FWHM decreases with decrease in temperature. The temperature dependence of FWHM can be expressed as [13]: $\Gamma(T) = \Gamma_0 + C(1+\frac{2}{e^x-1})$, where $\Gamma_0$ is full-width at half maxima at $T$ = 0 K, solid line in Figure 2(b) is the fitted curve using this equation with the fitted constant $\Gamma_0 = 3.94 \pm 0.10, C = 1.26 \pm 0.04$, in the temperature range of 60 K to 623 K.

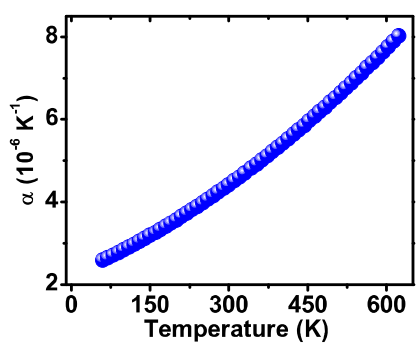

**FIGURE 3.** Temperature variation of thermal expansion coefficient of Si extracted from the Raman active phonon mode frequency.

## ACKNOWLEDGMENTS

The author, PK thanks Department of Science and Technology for the grant under INSPIRE faculty scheme.